\input psfig.tex
\documentstyle[11pt,paspconf]{article}
\newfont{\sss}{cmsy10 scaled 1200}

\begin{document}

\title{On the contribution of unresolved blazars to the extragalactic 
gamma ray background}

\author{A.~M\"ucke}
\affil{University of Adelaide, Adelaide, SA 5005, Australia}

\author{M.~Pohl}
\affil{Danish Space Research Institute, Juliane Maries Vej 30, 2100 K\o benhavn \O, 
Denmark}

\begin{abstract}
We present a model for estimating the blazar contribution to the
extragalactic $\gamma$-ray background (EGRB) in the EGRET energy range.
It is based on inverse-Compton scattering as the dominant
$\gamma$-ray production process in the jets of 
flat spectrum radio quasars (FSRQs) and BL~Lac objects, 
and on the unification scheme of radio-loud AGN.

\noindent Depending on the evolutionary behaviour of radio-loud AGN, our estimate 
for the point source contribution to the extragalactic background emission ranges
from about $\sim20\%-40\%$ for a redshift drop off at
$Z_{max}=3$ to around $\sim35\%-85\%$ for a cutoff at $Z_{max}=5$.
Roughly 70\%-90\% of the AGN contribution to the EGRB would be due to BL~Lacs/FR~I objects.
However, in case of BL Lacs our model parameters are not entirely
in agreement with the results from studies at radio wavelengths
indicating systematical uncertainties whereas our estimate for the FSRQ contribution
seems to be consistent. Thus we
end up with two possibilities depending on whether we underpredict or 
overpredict the BL Lac contribution: either unresolved AGN can not
account for the entire EGRB or unresolved BL Lacs produce the observed background.
\end{abstract}


\section{Introduction}

One of the most important results of the Energetic Gamma Ray
Experiment Telescope 
(EGRET) in the field of extragalactic astronomy is the measurement
of an isotropic diffuse $\gamma$-ray background over three decades of
photon energy (Sreekumar et al. 1998).
The differential spectrum can be described by a 
power law with photon spectral index $\alpha=2.10 \pm 0.03$.
The intensity of the EGRB integrated above 100 MeV is
$I=(1.45\pm0.05) \cdot 10^{-5}$ photons cm$^{-2}$ s$^{-1}$ sr$^{-1}$.

\noindent In this paper we follow a new ansatz for modeling the
blazar contribution to the EGRB, 
which is based on 
the unification scheme of radio-loud AGN and a particular
$\gamma$-ray emission model, namely external inverse-Compton
scattering of accretion disk photons (Dermer \& Schlickeiser 1993).

\section{The model}

According to the unification scheme of radio-loud AGN
blazars are the beamed sub-class of Fanaroff-Riley (FR) radio galaxies
(i.e. Urry \& Padovani 1995) with the most probable
parent population of FSRQs being the Fanaroff-Riley II (=FR~II) radio galaxies and
BL~Lacs being the beamed population of FR~I radio galaxies
(i.e. Padovani \& Urry 1992; Urry, Padovani \& Stickel 1991). 
For a homogeneous distribution of viewing angles and an
isotropic distribution of these objects throughout the universe, 
the $\gamma$-ray emission from radio-loud AGN observed under a large
viewing angle and/or at a large distance is expected to fall below the
EGRET-sensitivity limit. The 
integration of the $\gamma$-ray
intensity emitted by those objects
yields the point-source contribution to the EGRB.

\noindent We base 
the calculation of the non-thermal $\gamma$-ray emission of blazars
on the leptonic model of Dermer \& Schlickeiser (1993) and extend it
by considering not only instantaneous but also continuous electron
injection. In particular we consider a power law with index s=2 for the
injected electron spectrum and a fixed luminosity for the Shakura-Sunyaev-accretion
disk luminosity for all sources. The calculation of the template source
spectrum is described in detail in M\"ucke \& Pohl (1998).

\noindent 
Using a power law distribution for the energy injected into the jets of AGN,
$\rho_{FR}(E_{in}) = \frac{\rho_{0}}{10^{40}\rm{erg}} (\frac{E_{in}}{10^{40}\rm{erg}})^{-\delta}$ ($E_{in} \geq E_{in,1}$),
we calculate the time and energy integrated flux (from 100 MeV to
10 GeV) from BL~Lac/FR~I and FSRQ/FR~II objects.
The low energy cut-off E$_{in,1}$ and the normalization $\rho_0$ of the
distribution $\rho_{FR}(E_{in})$ are the free parameters of the model.

\noindent We use a Monte-Carlo method to
model the Log~N-Log~S and the redshift distribution 
of resolved BL~Lacs and FSRQs which are our model constraints.
%Our model is constrained by the Log~N-Log~S and redshift distributions 
%of the resolved BL~Lacs and FSRQs.

\noindent Objects which are brighter than the sensitivity limit determine the resolved Log~N-Log~S,
while sources with flux values below that limit contribute to the EGRB.
The unresolved blazar contribution to the EGRB then follows by summing up all unresolved
AGN per solid angle of the whole survey.
Our calculations consider a redshift drop off at $Z_{max}=3$ and $Z_{max}=5$.
Note that the $\gamma$-ray emission in the EGRET energy range is
moderated by pair production on the IR-background if it
originates at redshifts $Z>3$ (Stecker et al. 1992). The observed power
law spectrum of the EGRB up to 100 GeV therefore excludes significant 
contributions from sources at redshifts of $Z>3$.

\section{Results and conclusions}

The Log~N-Log~S and redshift distributions of both, FSRQs and BL~Lacs, were
compared with our model distributions by using a
Kolmogorov-Smirnov test. A compilation of the derived values
for the free parameters can be found in Table 1 for a redshift cutoff
at $Z_{max}=3$ and in M\"ucke \& Pohl (1998) for $Z_{max}=5$.
With the knowledge of E$_{in,1}$ and $\rho_0$ the distributions of the injected
energy of BL~Lacs/FR~I and FSRQs/FR~II can be constructed. In general
FR~I/BL~Lacs are more numerous than FSRQ/FR~II, but less luminous.
The total luminosity of all FR~I/BL~Lacs is similar to that of all
FSRQ/FR~II though.
As a consequence the background contribution of unresolved AGN is dominated by
FR~I/BL~Lacs ($\sim 70\%-80\%$), while resolved radio-loud AGN are mainly FSRQs.
With our model we can only explain $\simeq20\%-40\%$ of the observed extragalactic
$\gamma$-ray background as caused by unresolved radio-loud AGN. If the redshift
cut-off is assumed to occur at $Z_{max}=5$ this contribution increases
to $\simeq35\%-85\%$ of which $\simeq80\%-90\%$ would be BL~Lacs/FR~I.
The typical unresolved source in our model is either weakly beamed or has a weak intrinsic
luminosity of the blazar jet. 

\noindent 
In Figure 1 we show the predicted differential power
distribution $S^2 dN/dS$ of 
FSRQs/FR~II and BL~Lacs/FR~I for the case of
instantaneous electron injection and a source density cut-off at Z$_{max}=3$.
It indicates that the main contribution to the EGRB stems from
sources with a flux around
$S\sim10^{-8}$ ph cm$^{-2}$ s$^{-1}$. 
Our model predicts a significant flattening of the Log~N-Log~S distribution
in the next two decades of source flux below the EGRET threshold and number 
counts of roughly $\sim 0.1$ deg$^{-2}$.

\noindent Comparing the total number {\sss{N}} of BL~Lacs/FR~I and FSRQs/FR~II
from our model fit with the observations in the radio band 
(see Padovani \& Urry (1992),
Urry, Padovani \& Stickel (1991)) reveals a significant underestimate
of the total number of BL~Lac/FR~I-objects
while for the FSRQ/FR~II population our model is in agreement with the radio
observations. We note that the $\gamma$-ray emission model used in our study
may not be
ideally suited for BL Lacs anyway, since it is based on the
assumption that the accretion disk provides the target photons for the
dominant inverse-Compton process (EIC).
However, BL~Lacs
are thought to have rather weak accretion disks which may imply that
the SSC mechanism is the dominant $\gamma$-ray radiation process.
We therefore conclude that, depending  on whether we underpredict or 
overpredict the BL Lac contribution, {\it either unresolved AGN can not
account for the entire EGRB or unresolved BL Lacs/FR~I produce the observed background}.
The detection
of roughly $10^{2}-10^3$ more blazars in the energy range of $\sim 100$MeV - $10$GeV with a telescope $\sim 10$ times more 
sensitive compared to EGRET is expected.

\acknowledgments

AM is supported by a grant from the Australian Research Council.

%%%%%%%%%%%%%%%%%%%%%%% Z = 3 table %%%%%%%%%%%%%%%%%%%%%%%%%%%%%%%%%%%%%%%%%%%%%
\begin{table*}
\centering\scriptsize
\begin{tabular}{c|c|c|c|c}
& \multicolumn{2}{c}{\bf{FSRQs}} & \multicolumn{2}{|c}{\bf{BL~Lacs}} \\ \hline
& \bf{instan.inj.} & \bf{cont.inj.} & \bf{inst.inj.} & \bf{cont.inj.}\\ \hline
E$_{in,1}$ & $10^{44.4}$erg & $10^{44.4}$erg & $10^{43.8}$erg & $10^{43.8}$erg\\
$\rho_0$ & $(5.9\pm1.0)\cdot10^{6}$ & $(4.9\pm0.8)\cdot10^{6}$ & $80\pm27$ & $46\pm16$\\
{\sss{N}} & $(9.43\pm1.53)\cdot 10^4$ & $(7.82\pm1.26)\cdot 10^4$ & $(2.71\pm0.90)\cdot10^5$ & $(1.56\pm0.52)\cdot10^5$\\
{\sss{N}}[\%] & $26^{+5}_{-3}$ & $33^{+5}_{-3}$ & $74^{+3}_{-5}$ & $67^{+3}_{-5}$\\
P$_{NS}$ & 0.61 & 0.37 & 0.40 & 0.36\\
P$_{Z}$ & 0.65 & 0.71 & 0.96 & 0.78\\
$S_{unres}$ & $0.90\pm0.15$ & $0.99\pm0.16$ & $3.24\pm1.08$ & $2.78\pm0.93$\\
$S_{unres}$ [\%] & $6\pm1$ & $7\pm1$ & $22\pm7$ & $19^{+6}_{-7}$\\
$\frac{S_{unres}}{S_{unres,tot}}$ & $22^{+4}_{-2}$ & $26^{+5}_{-3}$ & $78^{+2}_{-4}$ & $74^{+3}_{-5}$\\
\end{tabular}
\caption[]{
The parameters of our model 
for $Z_{max}=3$ together with the
KS-probability for the statistical identity between the data and model
Log~N-Log~S distributions (P$_{NS}$) and redshift distributions (P$_{Z}$). 
E$_{in,1}$ and $\rho_0$ are the low energy cut-off and the normalization
of the injected energy distribution, respectively, and are the fit parameters
of our model. {\sss{N}} is the total number of objects of each object class
in the whole universe and $\rho_0$ (in Mpc$^{-3}$ erg$^{-1}$) the normalization of the distribution $\rho(E_{in})$.
$S_{unres}$ is the intensity (given in $10^{-6}$ photons cm$^{-2}$ s$^{-1}$ sr$^{-1}$ above 100 MeV and as fraction of the observed
background radiation) due to the sum of unresolved
BL~Lacs/FR~I and FSRQs/FR~II while $\frac{S_{unres}}{S_{unres,tot}}$ gives
the fractional contribution of each object class.
The given uncertainties are derived by varying the number of resolved objects
by $\pm\sqrt{N}$.
We predict that $\simeq20\%-40\%$ of the EGRB is due to unresolved AGN of which $\simeq70\%-80\%$ would be BL~Lacs/FR~I.}
\vspace*{-3cm}
\end{table*}
%%%%%%%%%%%%%%%%%%%%%%%%%%%%%%%%%%%%%%%%%%%%%%%%%%%%%%%%%%%%%%%%%%%%%%%%%%%%%%%%%%%%%%%%%%%%%%

\begin{figure}
%\centering
\begin{minipage}[t]{8.3cm}
\psfig{figure=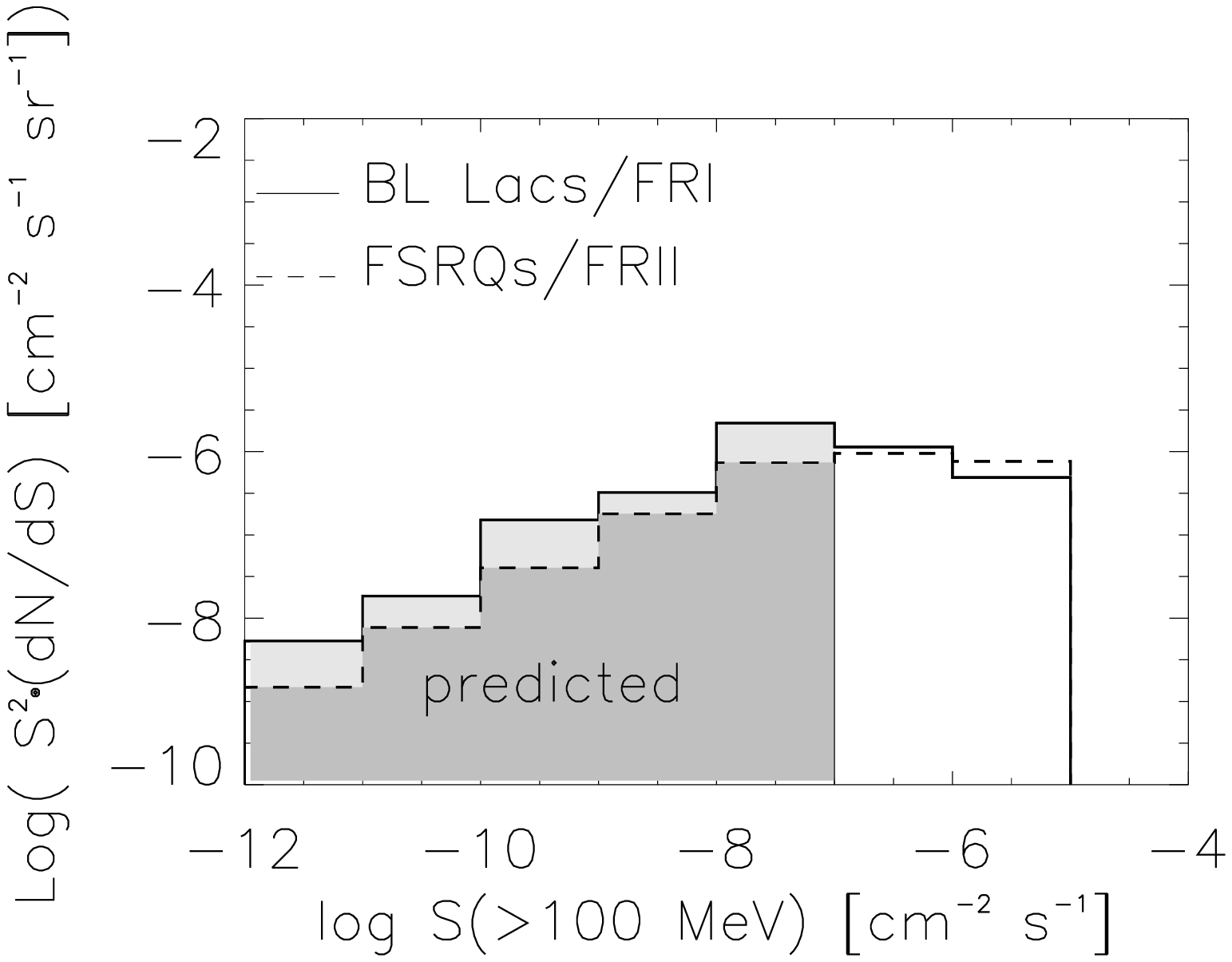,width=8.5cm,height=6.1cm,clip=}
\end{minipage}
\hfill
%\caption[]{
\begin{minipage}[t]{4.73cm}
\vspace*{-6cm}
Fig.1: Predicted Log~(S$^2$ (dN/dS))-Log~S-distribution of our background model
for BL~Lacs/FR~I and FSRQs/FR~II
for the case of a finite E$_{in}$-distribution and instantaneous electron
injection assuming a redshift cut off at
$Z=3$.
The statistical uncertainty in each bin lies at about 10\%.
The main contribution to the EGRB comes from
sources at $S\sim10^{-7 \ldots -8}$ phot cm$^{-2}$ s$^{-1}$.
\end{minipage}
%}
\end{figure}

\end{document}